\documentstyle[12pt,Dmath,titlepage]{article}
\begin{document}

  \def\thefootnote{\fnsymbol{footnote}}
%
  \title{
  \vspace*{-20mm}
  {\LARGE \bf 
  Hadley circulations \\
  and
  large scale motions of moist convection \\
  in the two dimensional numerical model
  \\[20mm]}}

  \author {{\Large \bf Masaki Satoh }\footnotemark[1] \\[5mm]
           {\Large Department of Mechanical Engineering,} \\
           {\Large Saitama Institute of Technology,} \\
           {\Large 1690 Fusaiji, Okabe, Saitama 369-02, Japan} \\[10mm]
          }
  \footnotetext[1]{{\it Corresponding author address}:\\
            Dr. Masaki Satoh, Department of Mechanical Engineering, 
            Saitama Institute of Technology, \\
            1690 Fusaiji, Okabe, Saitama 369-02, Japan\\
            E-mail: satoh@sit.ac.jp}

  \date   {7 Mar. 1994
           submitted to {\it J. Meteor. Soc. Japan}\\[5mm]
           5 Sep. 1995 revised}

  \maketitle

\setlength{\baselineskip}{8mm}

%
\begin{abstract}

As a tool for understanding
the meridional circulation of the atmosphere, 
a two-dimensional ( latitude -- height ) numerical model
is used to clarify
the relationship between the Hadley circulation
and large-scale motions associated with moist convection. 
The model is based on
the primitive equations including the moist process, 
and two kinds of coordinates are used:
the spherical coordinate
and the Cartesian coordinate with a uniform rotation. 
The surface temperature is externally fixed
and the troposphere is cooled by the radiation; 
unstable stratification generates
large-scale convective motions. 
\vspace{5mm}

Dependencies
on the surface temperature difference from north to south
$\Delta T_{s}$ are investigated. 
The numerical results show that 
a systematic multi-cell structure exists in every experiment. 
If the surface temperature is constant ($\Delta T_{s} = 0$ ),
convective motions are organized
in the scale of the Rossby deformation radius
and their precipitation patterns have
a periodicity of the advective time $\tau_{D}$. 
As $\Delta T_{s}$ becomes larger, 
the organized convective system
tends to propagate toward warmer regions. 
The convective cells
calculated in the Cartesian coordinate model
is very similar to those of the mid-latitudes
in the spherical coordinate model. 
In particular, 
the Hadley cell can be regarded 
as the limit of the convective cells
in the equatorial latitudes. 
\end{abstract}

  \pagenumbering{arabic}
  \def\thefootnote{\arabic{footnote}}

  \newpage
%
\section{Introduction}

  Although it is well understood that
  two-dimensional axisymmetric models
  are important for studying the Hadley circulation
  (Schneider, 1977; Held and Hou, 1980), 
  this kind of model has not been thought useful
  for large-scale motions in the extratropics, 
  since non-axisymmetric baroclinic waves are essential 
  in the real atmosphere. 
  Until the beginning of this century, however, 
  meridional cellar structure in an axisymmetric situation
  had been a major problem of the general circulation of the atmosphere
  (e.g., Thomson, 1892; Lorenz, 1967):
  not only the Hadley circulation
  but also the cellar structure in the extratropics. 
  Views of the general circulation has changed 
  from that given by Hadley(1735)
  in which only a direct cell exists in each hemisphere
  ( upward motion in the equatorial side
    and downward motion in the polar side)
  to that given by Ferrel(1856) and Thomson(1892)
  in which a direct cell co-exists with an indirect cell
  ( downward motion in the equatorial side
    and upward motion in the polar side). 
  Although various cellar structures had been presented, 
  the problem of the cell shapes was almost settled by 
  the requirement of the angular momentum balance; 
  the direct cell must override the indirect cell
  in the axisymmetric circulations, 
  and the cell boundary is inclined to the pole as in the upper layers. 
  This cellar structure is
  presented by Bjerknes(1937) and Eliassen(1951). 
  The numerical experiments by Schneider(1977) and Held and Hou(1980)
  have qualitatively reproduced this structure. 
  \vspace{5mm}

  Satoh(1994a) has calculated 
  the axisymmetric circulations of a moist atmosphere
  by using a two-dimensional model in the global domain
  (hereafter, referred to as S94). 
  The surface temperature is externally fixed
  and the free atmosphere is cooled by the gray radiation model. 
  Large-scale convective motions are naturally resolved in the model
  and the stratification is stabilized, 
  even if no cumulus parameterization except for
  the large-scale condensation is introduced. 
  The results show that, in the low-latitudes, 
  there exists a large-scale direct cell
  which corresponds to the Hadley cell. 
  The cell widths are broader as in the upper layers, 
  so that the cell shape is very similar to that presented
  by Bjerknes(1937) and Eliassen(1951). 
  In the extratropics, however, 
  unsteady multiple cells exist instead of an indirect cell
  and they propagate toward the equator
  (S94, Fig. 3;
   reproduced as Fig. 2(c) of the present paper). 
  Hereafter, the multiple cells calculated in S94
  in the mid- and high-latitudes
  are referred as the symmetric cells. 
  \vspace{5mm}

  The circulations in the mid-latitudes of S94
  are different from those obtained
  by Schneider(1977) and Held and Hou(1980). 
  The difference may be ascribed to the maintenance mechanism
  of the stratification of the models. 
  In S94, 
  the stratification tends to be unstable
  because of the radiative cooling in the troposphere 
  and the heating from the surface. 
  Convective motions must be generated
  in order to suppress the unstable stratification. 
  In contrast to this, 
  Schneider(1977) and Held and Hou(1980) used
  a stable temperature profile
  as a reference state of the Newtonian cooling, 
  so that no local convection exist in their models. 
  \vspace{5mm}

  In the extratropics of the real atmosphere, 
  non-axisymmetric motions such as baroclinic waves are dominant, 
  and the symmetric cells calculated in S94 is hardly observed. 
  As will be shown in sections 2 and 3, however, 
  the symmetric cells generally exist
  in all experiments of the two-dimensional model, 
  and play an important role in the transport of the angular momentum. 
  It is required that 
  dynamical mechanisms of the symmetric cells should be clarified, 
  in particular, for their propagation and horizontal scales, 
  to gain a complete view of the axisymmetric systems. 
  The symmetric cells are also valuable, 
  since they will provide a further insight into the Hadley circulation. 
  It will be shown that
  the Hadley cell can be regarded as one of the symmetric cells. 
  \vspace{5mm}

  In this paper, 
  the circulations in the mid-latitudes of the axisymmetric atmospheres
  are focused. 
  A two-dimensional model in horizontal-vertical section
  is utilized to investigate the dynamics of the symmetric cells. 
  Two kinds of coordinates are used; 
  the model in the spherical coordinate is referred
  as the global model, 
  and the model in the Cartesian coordinate is referred
  as the $f$--plane model. 
  Formulation of the model is described in the appendix. 
  In section 2, 
  the angular momentum budget of the standard experiment of S94 is analyzed, 
  and additional results for the global model are shown. 
  In sections 3 and 4, 
  the symmetric cells are discussed 
  with the results for the $f$--plane model.
  Section 5 summarizes
  the relationship between the symmetric cells and the Hadley circulations
  and discuss the generalization of these findings. 
  \vspace{5mm}

  \newpage
%
\section{Global condition}

\subsection{Angular momentum budget}

  The angular momentum budget plays a key role
  in meridional circulations of the axisymmetric atmosphere. 
  When steady direct and indirect cells co-exist
  as depicted by Bjerknes(1937) and Eliassen(1951), 
  the cell shapes are determined 
  by the requirements of the angular momentum balance
  in the following manner. 
  The direct cell transports the angular momentum upward, 
  and the indirect cell transports it downward, 
  so that there must be an exchange of the angular momentum
  between both cells. 
  In the axisymmetric system, 
  the viscous stress is the only process
  of the momentum exchange between steady cells. 
  If coefficients of the viscosity are isotropic, 
  horizontal transport caused by the viscous stress
  is much smaller than vertical transport by the viscous stress, 
  since the aspect ratio of the atmosphere is very small. 
  Thus, the angular momentum is 
  vertically transported by the viscous stress
  which is associated with the vertical shear of the zonal winds. 
  In order for the zonal shear to be maintained, 
  the direct cell must override the indirect cell, 
  and the cell boundary is inclined to the pole as in the upper layers. 
  \vspace{5mm}

  In the beginning, 
  we examine the angular momentum budget 
  of the results for the global model of S94, 
  and compare it with the above consideration. 
  In the standard experiment of S94, 
  the surface temperature is given by
  \begin{equation}
  T_{s}(\varphi) 
  = T_{A} - ( T_{B} - T_{A} ) \sin^{2} \varphi, 
                                                    \label{eq:Ts}
  \end{equation}
  where $\varphi$ is the latitude, 
  $T_{A} = 300$ K, and $T_{B} = 260$ K. 
  The diffusion coefficient of momentum flux is 5 m$^{2}$/s, 
  and that of heat and moisture fluxes is 1 m$^{2}$/s. 
  \vspace{5mm}

  In the standard experiment, 
  the Hadley cell exists in the low-latitudes, 
  and a multiple cellar structure which propagates equatorward 
  exists in the mid- and high-latitudes. 
  In order to isolate transports 
  by the symmetric cells in the mid- and high-latitudes, 
  we divide convergence of the angular momentum
  into steady and unsteady parts
  \footnotemark:
  \footnotetext{
    In precisely, 
    the convergence is defined by
    $- \Ddiv ( p_{s} l \Dvect{v} ) / p_{s}$, 
    since the model is expressed by the $\sigma$ coordinate. 
    Fig. 1(a) is the total convergence
    $- \overline{\Ddiv ( p_{s} l \Dvect{v} ) / p_{s}}$, 
    (b) is the steady part
        $- \Ddiv ( \overline{p_{s}} \overline{l} \overline{\Dvect{v}} )
         / \overline{p_{s}}$, 
    and (c) is the difference between (a) and (b). 
  }
  \begin{equation}
  - \overline{ \Ddiv ( l \Dvect{v} )}
  = - \Ddiv ( \overline{l} \overline{\Dvect{v}} )
  - \Ddiv ( \overline{l' \Dvect{v'}} ). 
                                                    \label{eq:lcnv}
  \end{equation}
  $l$ is the absolute angular momentum, 
  $\Dvect{v}$ is the meridional velocity, 
  and $\Ddiv{}$ is the divergence in the meridional plane. 
  $\overline{()}$ denotes time average
  and $()'$ the deviation from the average. 
  Fig. 1 shows the respective terms in Eq. (\ref{eq:lcnv}). 
  The total component (a) shows a large divergence
  ( gain of the angular momentum in the atmosphere )
  in the lower boundary layers of the Hadley cell. 
  At the polar boundary of the Hadley cell 
  (the latitudes 10$^{\circ}$--20$^{\circ}$), 
  divergent and convergent regions
  are vertically adjacent to each other. 
  Both regions are inclined to higher latitudes in the upper layers. 
  The steady component (b)
  is uniformly divergent in the lower layers in every latitude, 
  while it is convergent in the upper layers 
  in the extratropics. 
  Conversely, 
  the unsteady component (c) is divergent in the upper layers, 
  while it is convergent in the lower layers. 
  \vspace{5mm}

  Fig. 1 indicates that
  it is the symmetric cells 
  that transport the angular momentum downward in the mid-latitudes. 
  The angular momentum is supplied to the atmosphere
  in the surface easterly regions in the low-latitudes, 
  and is transported upward and poleward by the Hadley cell. 
  It is transferred downward by the viscous stress
  at the polar boundary of the Hadley cell, 
  where a large vertical shear of the westerlies exists. 
  In the mid- and high-latitudes, 
  westerly shear is maintained
  by the poleward transport of the angular momentum
  by a weak steady direct cell, 
  which has a hemispheric scale as shown by Fig. 1(b). 
  On the other hand, 
  the unsteady symmetric cells transport the angular momentum downward
  in these regions. 
  Although both easterlies and westerlies exist
  at the surface boundary of the symmetric cells, 
  the westerly component is a little stronger, 
  hence the angular momentum is transferred
  from the atmosphere to the ground
  due to the net effect of the surface friction. 
  The role of the Hadley cell in the angular momentum budget
  is the same as that depicted by Bjerknes(1937) and Eliassen(1951). 
  In the extratropics, however, 
  the symmetric cells play a role of the downward transport
  of the angular momentum
  instead of a steady indirect cell. 
  \vspace{5mm}

\subsection{Additional experiments}

  Additional experiments of the global model are performed, 
  so as to investigate the dependencies of the symmetric cells
  on the surface temperature differences 
  $\Delta T_{s} = T_{A} - T_{B}$
  and the viscosity coefficients $\nu$. 
  Fig. 2 shows
  the time sequences of the latitudinal distribution of the precipitation
  for $\Delta T_{s} = $ 0K, 20K, 40K (the standard experiment), and 80K. 
  The temperature at the equator is $T_{A} =$ 300K for every case. 
  In the standard experiment, 
  the symmetric cells are propagating equatorward
  at a speed of approximately 1 m/s, 
  and have a period of about 20 days. 
  The speed of the propagation becomes faster
  as $\Delta T_{s}$ becomes larger. 
  In the case of $\Delta T_{s} = 0$, 
  the precipitation patterns are periodic
  and they do not change their positions
  except for the equatorial latitudes. 
  These results imply that
  the gradient of the surface temperature
  and the Coriolis parameter affect
  the propagation of the symmetric cells. 
  It should be noted that, 
  in the case of $\Delta T_{s} =$ 0 (Fig. 2(a)), 
  precipitation is not concentrated at the equator
  but randomly distributed in the low-latitudes. 
  That is, the Hadley cell no longer exists in this model, 
  when the surface temperature gradients 
  are sufficiently small near the equator. 
  \vspace{5mm}

  Fig. 3 displays 
  the dependencies of the precipitation 
  on the viscosity coefficients of the free atmosphere ( $\nu$ )
  and of the surface layer ( $\nu_{s}$ ):
  $\nu = \nu_{s} =$ 1 and 25 m$^{2}$/s. 
  In the case of $\nu = \nu_{s} =$ 25 m$^{2}$/s, 
  the symmetric cells around the latitudes 30$^{\circ}$
  do not propagate. 
  This result indicates
  the propagation of the symmetric cells
  is related with the angular momentum transport, 
  since the viscous diffusion controls the vertical momentum transport. 
  \vspace{5mm}

  \newpage
%
\section{Mid-latitude condition}

\subsection{Model}

  In this section,
  an "$f$-plane model" is used
  to investigate the symmetric cells in the mid-latitudes; 
  the model is two-dimensional (horizontal-vertical)
  with a uniform rotation rate in the Cartesian coordinate. 
  As summarized in the appendix, 
  the formulation of the model, 
  i.e. the basic equations, the discretization, 
  and the physical processes, 
  is almost the same as that of the global model. 
  Only the metrics and the domain are different. 
  \vspace{5mm}

  The model is based on the primitive equations
  with a $\sigma$ coordinate
  ( $\sigma = p / p_{s}$; 
  $p$ the pressure and $p_{s}$ the surface pressure ). 
  The lateral widths of the domain is $y_{0} = 10^{7}$ m. 
  Here, the lateral coordinate is denoted by $y$, 
  which corresponds to the latitude of the global model;
  larger $y$ indicates the northward direction. 
  The lateral boundaries are assumed to be walls
  with slip and insulating conditions. 
  The number of grid points is 100 in the horizontal direction
  and 50 in the vertical direction. 
  The grid intervals are $\Delta y/y_{0} =$ 0.02
  and $\Delta \sigma =$ 0.02, respectively. 
  The Coriolis parameter is prescribed as
  $f = \Omega_{0} = 7.272 \times 10^{-5}$ rad/s, 
  which is the value at the latitude 30$^{\circ}$
  ($\Omega_{0}$ is the rotation rate of the earth ).
  \vspace{5mm}

  As for the radiation process, 
  a uniform cooling rate 2K/day is given
  below the level $\sigma = 0.1$ to the ground just for simplicity. 
  The level $\sigma = 0.1$ can be regarded as the tropopause. 
  The surface temperature is externally fixed, 
  and the ground surface is assumed to be wet everywhere. 
  Moisture is supplied from the surface to the atmosphere 
  by the bulk formula. 
  In the same way as section 3a of S94, 
  only the ``large-scale condensation''
  is employed as the moist process in the atmosphere, 
  and any other cumulus parameterization is used; 
  phase change of moisture is explicitly calculated
  at each of the grid point, 
  and the liquid part of the air is removed as precipitation
  when it is saturated. 
  Diffusion coefficients for the momentum, heat, and moisture
  are set to be constant as in S94. 
  \vspace{5mm}

\subsection{Results}

  The surface temperature is linearly prescribed as
  \begin{equation}
  T_{s} = T_{A} + ( T_{A} - T_{B} ) \frac{y}{y_{0}}. 
                                                    \label{eq:Ts_f}
  \end{equation}
  The temperature difference in the domain is denoted
  by $\Delta T_{s} = T_{A} - T_{B}$. 
  Fig. 4 shows
  time variations (0 -- 100 days) 
  of precipitation for
  (a) $T_{A} = T_{B} =$ 300 K,
  (b) $T_{A} =$ 305 K, $T_{B} =$ 295 K, 
  (c) $T_{A} =$ 310 K, $T_{B} =$ 290 K, 
  and (d) $T_{A} =$ 320 K, $T_{B} =$ 280 K. 
  In every case, 
  there exists a systematic cellar structure 
  similar to that in the mid-latitudes of the global model. 
  The precipitation occurs 
  in the upward motion regions of the symmetric cells. 
  When there is no temperature difference ( $\Delta T_{s} =$ 0 ), 
  the precipitation patterns have a periodicity of about 20 days. 
  When there is a temperature difference ( $\Delta T_{s} \neq$ 0 ), 
  the precipitation patterns propagate to the warmer regions
  ( smaller $y$ ). 
  As the temperature difference becomes larger, 
  the velocity of the propagation becomes larger. 
  Although the lateral intervals between the precipitating regions
  are not sensitive to $\Delta T_{s}$, 
  they are larger in the warmer regions. 
  The periodicity or the propagation of the precipitation
  is associated with the circulations of the symmetric cells. 
  Structure and mechanism of the symmetric cells
  will be examined in the following subsections. 
  \vspace{5mm}

\subsection{Case of $\Delta T_{s} =$ 0}

  The symmetric cells in the case $\Delta T_{s} =$ 0 
  (Fig. 4(a)) are examined here. 
  Fig. 5 shows the time variation of the surface zonal winds
  for the days 0 -- 100. 
  \footnotemark
  \footnotetext{
  The winds perpendicular to the model domain refer to the zonal winds; 
  positive winds are westerly, 
  and negative winds are easterly. }
  Westerlies exist in the southern side of the precipitating regions, 
  whereas easterlies exist in the northern side. 
  Their phases correspond to those of the precipitation. 
  Figs. 6(a) and (b) show
  the zonal winds in the meridional plane ($y$-$\sigma$)
  for the averages
  of the days 46--55 and the days 56--65, respectively. 
  Figs. 7(a) and (b) are the temperature perturbation
  from the $y$-average for the same periods. 
  The phase of the precipitation changes at about the day 55. 
  For instance, 
  precipitation occurs at $y/y_{0} =$ 0.52 during the days 46--55, 
  and it jumps to $y/y_{0} =$ 0.41 and 0.62 at the day 55. 
  At the precipitating regions, 
  the circulations of the lower layers
  are cyclonic with convergence of lateral winds, 
  while those of the upper layers
  are anticyclonic with divergence of lateral winds. 
  The temperatures in the upward motion regions are warmer
  than those in the downward motion regions
  irrespective of height. 
  The temperatures are almost
  in thermal wind balance with the zonal winds. 
  \vspace{5mm}

  The periodicity of the symmetric cells can be explained as follows. 
  Associated with the convective circulation, 
  cyclonic winds develop in the lower layers, 
  and anticyclonic winds develop in the upper layers. 
  In the northern part of a precipitating region
  (upward motion region), 
  easterlies develop in the lower layers
  while westerlies develop in the upper layers. 
  As the convective motion continues, 
  the westerlies in the upper layers gradually come down. 
  By the time they reach the ground, 
  the lateral gradient of pressure changes its sign
  due to the geostrophic adjustment to the westerlies. 
  The inverted pressure gradient force
  causes lateral winds of the opposite direction near the ground. 
  Finally, 
  the circulations are reversed;
  the downward motions are generated 
  in the previously upward motion region, 
  and upward motions are generated 
  in the previously downward motion region. 
  The period is given by the overturning time $\tau_{D}$, 
  which is defined by $H/w_{d}$
  where $H$ is the height of the troposphere
  and $w^{d}$ is the speed of the downward motion. 
  (This is also estimated by Eq. (29) of S94. )
  \vspace{5mm}

  The lateral scale of the symmetric cells $L$
  can be estimated in the following manner. 
  Note that 
  there are two cells between adjacent precipitating regions, 
  so that the lateral interval of the precipitating regions
  is expressed by $2L$. 
  If one denotes the lateral temperature difference
  in the middle of the troposphere by $\Delta T$, 
  the thermal wind balance is expressed by
  \begin{equation}
  \frac{g}{T_{0}}\frac{\Delta T}{L} = \frac{fU/2}{H}, 
                                                    \label{eq:dtdy_f}
  \end{equation}
  where $U$ is the maximum value of the zonal wind
  at the tropopause $H$, 
  and $T_{0}$ is the mean temperature. 
  By estimating $U = fL$
  from the angular momentum conservation\footnotemark 
  at the height $H$, 
  \footnotetext{
     The angular momentum in the $f$--plane
     is expressed by $u - fy$. 
  }
  one obtains
  \begin{equation}
  L = \left( \frac{2gH}{f^{2}} \frac{\Delta T}{T_{0}} \right)^{1/2}, 
                                                    \label{eq:L_f}
  \end{equation}
  where $g$ is the acceleration of gravity. 
  \vspace{5mm}

  The value of $\Delta T$ should be known in order to obtain $L$. 
  As shown below, 
  only an upper bound of $\Delta T$ can be given
  from the consideration of the stratification. 
  However, it will be shown that
  $\Delta T$ is actually close to the upper bound. 
  If one supposes a thermal wind balance of a large $L$, 
  $\Delta T$ also becomes larger as $L$ is increased. 
  In this case, 
  the temperature profile in the upward motion regions does not change, 
  since it is prescribed by a moist adiabat. 
  Thus, 
  the temperature in the downward motion regions must be cooler
  in this balance. 
  Since the surface temperature is uniform, 
  the stratification of the downward motion regions becomes unstable. 
  This implies that
  another convective motion will occur from these regions. 
  Although the downward motion regions are, 
  in general, conditionally unstable, 
  they would be absolutely unstable 
  in the balance of a sufficiently large $L$. 
  Therefore, 
  the upper bound of $\Delta T$ is given by that in the case
  when temperature lapse rate in the downward motion regions
  is equal to the dry adiabat. 
  Fig. 8 displays a time variation of lapse rate
  ( $-dT/dz$ )
  at $\sigma = 0.9$ which is just above the mixed layers. 
  The values above 8 K/km are shown by hatching. 
  Comparison with Fig. 4(a) shows that
  the lapse rate is about 9K/km
  at the downward motion regions
  when the circulation changes its direction. 
  This lapse rate is close to the dry adiabat 
  $\Gamma_{d} = g / C_{p} =$ 9.8K/km
  ( $C_{p}$ is specific heat for constant pressure ). 
  From this, 
  one may conclude that
  the temperature difference $\Delta T$ is given by this upper bound.
  \vspace{5mm}

  The temperature difference is given by
  \begin{equation}
  \Delta T = \left( T_{s} - \Gamma_{m} \frac{H}{2} \right)
           - \left( T_{s} - \Gamma_{d} \frac{H}{2} \right)
           = \left( \frac{g}{C_{p}} - \Gamma_{m} \right) \frac{H}{2}, 
                                                    \label{eq:DeltaT_f}    
  \end{equation}
  where $\Gamma_{m}$ is the lapse rate of the moist adiabat. 
  One obtains, therefore, 
  by using Eq. (\ref{eq:L_f}) and (\ref{eq:DeltaT_f}), 
  \begin{equation}
  L = \left[ \frac{H^{2}}{f^{2}} 
             \frac{g}{T_{0}} 
             \left( \frac{g}{C_{p}} - \Gamma_{m} \right) \right]^{1/2}
    = \frac{NH}{f}
    \equiv R_{N}, 
                                                    \label{eq:L_f2}
  \end{equation}
  where $N^{2} = g / T_{0} ( g / C_{p} - \Gamma_{m} )$
  is the Brunt V\"{a}is\"{a}l\"{a} frequency 
  for the moist adiabatic profile. 
  In conclusion, 
  the lateral scale 
  is given by the Rossby deformation radius $NH/f$. 
  It is estimated as
  $L \approx 0.11 y_{0} = 1.1 \times 10^{6}$
  for the values in the case of Fig. 4(a): 
  $N^{2} = 7 \times 10^{-5}$ [sec$^{-2}$], 
  $H = 1 \times 10^{4}$ [m] and
  $f = 7.27 \times 10^{-5}$ [sec$^{-1}$]. 
  Fig. 4(a) shows that the lateral interval of the precipitating regions
  is approximately 0.18 $y_{0}$;
  that is, $L \approx 0.09 y_{0}$, 
  which is very close to the above estimation. 
  \vspace{5mm}

  One might suspect that
  the result that the scale of the convective cell is given 
  by the Rossby deformation radius
  is only a trace of linear theories
  (e.g., Gill, 1982; Emanuel, 1983; Bretherton, 1987). 
  However, 
  the present case is considered
  in the nonlinear statistically equilibrium states, 
  to which the linear theories are not directly applied. 
  The nonlinear advection of the angular momentum
  and the asymmetry
  between the lower and the upper boundary conditions
  for the momentum
  (the angular momentum is exchangeable at the lower surface, 
  but it is not at the upper boundary)
  are key points of the above derivation. 
  If, instead, 
  one applies the slip conditions in which
  momentum exchange is forbidden
  both at the lower and the upper boundaries, 
  the angular momentum will be horizontally uniform 
  in the whole layer around the cumulus region
  in the nonlinear equilibrium states. 
  In this case,  
  the vertical component of the absolute vorticity becomes zero, 
  so that the circulation field associated with cumulus heating 
  will not feel the basic rotation $f$
  and the scale of the convective cell
  will be larger than the Rossby deformation radius. 
  In the present study, however, 
  the circulation field can feel 
  the basic rotation
  through the angular momentum exchange at the lower surface, 
  and, as a result, 
  the Rossby deformation radius 
  becomes a natural scale for the convective cell. 
  \vspace{5mm}

\subsection{Case of $\Delta T_{s} \neq$ 0}

  As an example for $\Delta T_{s} \neq$ 0, 
  the wind fields in the case of $T_{A} =$ 320 K and $T_{B} =$ 300 K
  are examined. 
  Fig. 9 shows a time variation (the days 0 -- 100) of the precipitation. 
  \footnotemark
  \footnotetext{
    As for Figs. 4(c) and 9, 
    $\Delta T_{s} =$ 20K is the same, 
    but $T_{A}$ and $T_{B}$ are different. 
    These figures show that
    the absolute values of temperature
    affect the lateral intervals of the precipitating regions; 
    the intervals are larger as the temperature is higher. 
  }
  The wind fields are averaged for 50 days
  along the propagating precipitating region
  A $\rightarrow$ B of Fig. 9
  ( A : $y/y_{0} =$ 0.68 at the day 51, 
    B : $y/y_{0} =$ 0.45 at the day 100 ). 
  Figs. 10 (a), (b) show
  the zonal winds and the stream functions, respectively. 
  The line A $\rightarrow$ B corresponds to $y/y_{0} =$ 0.5 of Fig. 10. 
  In the lower layers, 
  the northern side of the precipitating region ( $y/y_{0} =$ 0.5 )
  is westerly, 
  while the southern side is easterly. 
  In the upper layers, contrary to this, 
  the northern side of $y/y_{0} =$ 0.5 is easterly,
  while the southern side is westerly.
  The region of the upper westerlies
  is connected with that of the surface westerlies
  ( $y/y_{0} =$ 0.6 -- 0.66 ), 
  which are associated with the inflow
  to the next precipitating region
  ( $y/y_{0} =$ 0.66 ). 
  As shown by the stream functions, 
  each precipitating region has 
  both northern and southern cells. 
  The northern cell is broader than the southern cell
  especially in the upper layers, 
  and the northern cell spreads over 
  the next southern cell
  which is associated 
  with the next precipitating region ( $y/y_{0} =$ 0.66 ). 
  That is, 
  the direct cell ( positive sign of the stream functions )
  is in contact with
  the indirect cell ( negative sign )
  by an inclined boundary ( the contour of zero ). 
  The angular momentum is transported upward by the direct cell, 
  and is transported downward by the indirect cell, 
  hence the angular momentum should be transferred 
  from the direct cell to the indirect cell. 
  The inclination of the cell boundary
  makes possible
  the vertical transfer between the cells by the viscosity. 
  This cellar structure is similar to the relation
  between the Hadley cell and the Ferrel cell depicted by Eliassen(1951), 
  and is essential
  for the angular momentums transport between the two cells. 
  \vspace{5mm}

  Fig. 11 shows the temperature 
  and the normalized angular momentum $(u-fy)/(fy_{0})$
  at the day 100. 
  The upper angular momentum and temperature is uniform
  in the regions $y/y_{0} =$ 0.22 -- 0.4
  and $y/y_{0} =$ 0.41 -- 0.64, for example. 
  These uniform regions are associated with 
  the precipitation at $y/y_{0} =$ 0.24 and  0.46, respectively
  (Fig. 9). 
  The lateral gradient of the temperature
  is larger as in the lower layer, 
  and the temperature is in the thermal wind balance. 
  In this respect, 
  the structure of the symmetric cells is very similar 
  to that of the Hadley cell; 
  the angular momentum is conserving in the upper layers
  and the temperature is balanced with the vertical shear
  (compare with Figs. 2 and 4 of S94). 
  \vspace{5mm}

  As for the relationship with the Hadley cell, 
  the symmetric cells in the $f$-plane model should be investigated
  in terms of the argument by Plumb and Hou(1992), 
  in which the symmetric circulations are classified 
  into two regimes: 
  a non-meridional circulation regime
  in thermal wind balance with the prescribed heating
  ( a thermal equilibrium (TE) solution)
  and a meridional circulation regime
  with an angular momentum conserving flow in the upper layer
  ( an AMC solution ). 
  They argued that
  the response takes the form of the TE solution
  if the thermal forcing is weak or broad
  or it is far from the equator
  (Eq. (8) and Appendix of Plumb and Hou). 
  Although the Newtonian cooling is used as the thermal forcing
  in Plumb and Hou, 
  the similar argument is partly applied to the present study 
  if their reference temperature $\hat{T}_{e}$ 
  is replaced by the surface temperature $T_{s}$. 
  In the $f$-plane model, 
  Eq. (8) of Plumb and Hou corresponds to
  \begin{equation}
    - \frac{gH}{T_{0}} \DP[2]{T_{s}}{y} < f. 
                                                    \label{eq:TE}
  \end{equation}
  If this condition is satisfied, 
  the angular momentum balanced with the surface temperature
  does not have an extremum
  in the interior of the atmosphere
  and the TE solution is regular provided. 
  Plumb and Hou further argued that
  the AMC solution cannot exist
  when the TE solution is regular provided. 
  If the surface temperature is linearly distributed
  (Eq. (\ref{eq:Ts_f})), 
  the left hand side of (\ref{eq:TE}) vanishes
  and the inequality holds. 
  However, 
  it should be pointed out that
  the proof in their Appendix is insufficient for the $f$-plane model, 
  since they implicitly assumed
  that two Hadley cells coexist: the summer cell and the winter cell. 
  Their relation (A13), which is derived from this assumption, 
  does not hold in the present situation. 
  Thus, the possibility of the AMC solution is not precluded
  even if (\ref{eq:TE}) is satisfied. 
  \vspace{5mm}

  In contrast to Plumb and Hou, 
  S94 derived the condition for the AMC solution
  and the latitudinal scale of the Hadley cell
  from the relation between the surface temperature $T_{s}$
  and the temperature in the middle layer $T$
  which is in the thermal wind balance
  with the angular momentum conserving flow
  ( Figs. 11 and 13 of S94 ). 
  From this, 
  the AMC solution can exist
  when $T$ is higher than $T_{s}$ around the upward motion region. 
  This condition is rewritten in terms of the gradients:
  \begin{equation}
    \left| \DP{T_{s}}{y} \right| > \left| \DP{T}{y} \right|. 
                                                    \label{eq:AMC}
  \end{equation}
  The temperature deviation in the middle layer
  from that at the upward motion region of the AMC solution 
  is given by $\Delta T$ in Eq. (\ref{eq:L_f}). 
  Replacing $L$ by the distance from the upward motion region
  $\delta y$
  and $\Delta T$ by $- \delta T(\delta y)$, 
  one obtains
  \begin{equation}
  \delta T(\delta y)
  = - \frac{T_{0} f^{2} \delta y^{2}}{2 g H}. 
                                                    \label{eq:T_AMC}
  \end{equation}
  From this, 
  the right hand side of (\ref{eq:AMC}) 
  becomes zero as $\delta y \rightarrow$ 0, 
  so that the AMC solution can be possible 
  if there is a non-zero surface temperature gradient. 
  If one denotes the deviation of $T_{s}$
  from that at the upward motion region
  by $\delta T_{s}(\delta y)$, 
  the relation between $\delta T_{s}$ and $\delta T$
  gives the scale of the cell $L$ as shown by Fig. 12(b). 
  By equating the change of $\delta T$ in $L$
  with that of $\delta T_{s}$:
  \begin{equation}
  \delta T (L) = \delta T_{s}(L)
               = - \Delta T_{s} \frac{L}{y_{0}}, 
                                                    \label{eq:DeltaT_f2}
  \end{equation}
  one obtains with Eq. (\ref{eq:T_AMC}), 
  \begin{equation}
  L = \frac{2gH \Delta T_{s}}{f^{2} y_{0} T_{0}} \equiv R_{T}. 
                                                    \label{eq:L_delT_f}
  \end{equation}
  This indicates that $L$ increases with $\Delta T_{s}$. 
  \vspace{5mm}

  The scale of the cell $L$ corresponds to the cell width of
  the northern side of the precipitating region; 
  the cell width is defined
  by the upper spread of the stream functions of the cell. 
  (Remind that $L$ is not the interval of the precipitating regions. 
  In the case of $\Delta T_{s} =$ 0, 
  the cell width is $L$ 
  whereas the interval of the precipitating regions is $2L$. )
  For $\Delta T_{s} = 20$ K and $T_{0} =$ 270K, 
  Eq. (\ref{eq:L_delT_f}) estimates $L = 0.027 y_{0}$, 
  which is much smaller than the cell width shown by Fig. 10. 
  Furthermore, 
  Eq. (\ref{eq:L_delT_f}) approaches zero
  as $\Delta T_{s} \rightarrow 0$
  and contradicts Eq. (\ref{eq:L_f2}). 
  One of the reasons for this underestimation is that
  the effect of static stability, 
  which is important in the case of $\Delta T_{s} =$ 0, 
  is not taken into account
  in the derivation of Eq. (\ref{eq:L_delT_f}). 
  As shown by Fig. 11, 
  the temperatures in the middle layers are maximum 
  around $y/y_{0} = 0.46$ where the precipitation occurs, 
  and the temperatures are minimum near $y/y_{0} =$ 0.40 and 0.64. 
  The northern and the southern cells
  exist in the regions between these minimum and maximum temperatures;
  the northern cell spreads in the region $y/y_{0} =$ 0.46 -- 0.64, 
  and the southern cell spreads in the region $y/y_{0} =$ 0.40 -- 0.46. 
  As in the case of $\Delta T_{s} =$ 0, 
  the static stability is smaller around the minimum temperatures. 
  One can take the effect of the static stability on the cell scale
  into account
  by simply adding Eq. (\ref{eq:DeltaT_f}) to Eq. (\ref{eq:DeltaT_f2})
  (Fig. 12(c)):
  \begin{equation}
  \delta T (L) = - \Delta T_{s} \frac{L}{y_{0}} 
               - \left( \frac{g}{C_{p}} - \Gamma_{m} \right) \frac{H}{2}, 
                                                    \label{eq:DeltaT_f3}
  \end{equation}
  which yields with Eq. (\ref{eq:T_AMC}), 
  \begin{equation}
  L^{2} - R_{T} L - R_{N}^{2} = 0. 
                                                      \label{eq:L3_relation}
  \end{equation}
  From this, one obtains
  \begin{equation}
  L_{\pm} = \frac{R_{T}}{2} \pm \sqrt{ \frac{R_{T}^{2}}{4} + R_{N}^{2} }, 
                                                      \label{eq:L3}
  \end{equation}
  where $L_{+}$ is the width of the northern cell
  and $- L_{-} (> 0)$ is the width of the southern cell. 
  If one notes
  \begin{equation}
  \frac{R_{T}}{R_{N}} = \frac{2 g \Delta T_{s}}{f N y_{0} T_{0}}
                      \approx 0.24 \frac{\Delta T_{s}}{20\mbox{K}}
                      \ll 1, 
                                                      \label{eq:RTRN}
  \end{equation}
  then the interval of the precipitating regions is given by
  \begin{eqnarray}
  L_{+} - L_{-} &=& 2 \sqrt{ \frac{R_{T}^{2}}{4} + R_{N}^{2} }
                                                      \nonumber \\
    &\approx& 2 R_{N} \left( 1 + \frac{R_{T}^{2}}{8 R_{N}^{2}} \right)
    \approx 2 R_{N} \left[ 1 + 0.007 
      \left( \frac{\Delta T_{s}}{20\mbox{K}} \right)^{2} \right], 
                                                   \label{eq:L3_approx}
  \end{eqnarray}
  which is not sensitive to $\Delta T_{s}$. 
  This result is consistent with Fig. 4, 
  in which the intervals of the precipitating regions
  are almost the same regardless of $\Delta T_{s}$:
  0.18 $y_{0}$ in the case of $\Delta T_{s} =$ 0
  and 0.2 $y_{0}$ in the case of $\Delta T_{s} =$ 40K. 
  \vspace{5mm}

  The reason for the propagation of the symmetric cells
  is suggested as follows. 
  In the case of $\Delta T_{s} =$ 0, 
  as shown in the previous subsection, 
  the symmetric cells are periodic rather than propagating. 
  It has been argued that, 
  if the upper angular momentum descends in the northern side, 
  zonal winds become westerlies, 
  and that 
  these westerlies lead to the change of direction of the circulation
  and hence the periodicity. 
  In order for the symmetric cells 
  to preserve their circulation structure, 
  the surface winds must be easterly
  in the northern side of the precipitating region. 
  The zonal winds can be easterly
  only if the symmetric cells propagate southward
  during the overturning period, 
  since the upper angular momentum descends to a more southern place
  than the place where it is supplied
  (this is schematically shown by Fig. 13). 
  The above consideration suggests that, 
  as a convenient estimation, 
  the upper bound of the speed of the propagation 
  of the symmetric cells is given by $L/\tau_{D}$, 
  where $\tau_{D}$ is the overturning time and $L$ is the cell length. 
  This corresponds to the ultimate case 
  in which the angular momentum is conserved (Fig. 13). 
  The value of the angular momentum is reduced
  by the viscous transfer
  with the adjacent indirect cell in the northern side, 
  and it is also preferable for the easterly. 
  The viscosity also affects the speed as shown by Fig. 3, 
  though this effect is not easy to be taken into account
  in the simple scaling argument. 
  \vspace{5mm}

  It should be noted, however, 
  that the above consideration does not tell
  which is the preferable direction for the propagation. 
  The experiments show that 
  the symmetric cells always propagate southward
  (toward the warmer regions). 
  \footnotemark
  \footnotetext{
    Even in the case of $\Delta T_{s} =$ 0,
    the symmetric cells occasionally propagate. 
    For instance,
    Fig. 4(a) shows that there is a precipitating region
    propagating toward larger $y$ near $y \approx$ 0.9
    during the days 20 -- 35.
    This propagation seems to occur by chance. 
  }
  The direction of the propagation seems to be controled
  by the requirement of the surface easterlies. 
  One can suppose a basic state
  that has a westerly shear in thermal wind balance 
  with the prescribed surface temperature gradient. 
  \footnotemark
  \footnotetext{
     In fact, the westerly shear is produced
     by the symmetric cells themselves. 
     However, 
     the basic state is formally defined
     by the steady component of the zonal winds. 
  }
  The upper westerly does not change its direction
  if it descends at the same $y$. 
  Only if it descends southward, it becomes easterly. 
  There must be both easterlies and westerlies
  in the surface zonal winds
  on the constraint of the angular momentum conservation. 
  The symmetric cells should move southward as a whole
  in order to cancel the basic westerly component
  and to produce surface easterlies by the advection. 
  If they move northward, 
  westerlies would be stronger at the surface. 
  \vspace{5mm}

  \newpage
%
\section{Relation with the symmetric instability}

In this section, 
the symmetric cells in the case of $\Delta T_{s} \neq$ 0
are examined in terms of the symmetric instability (SI)
(e.g., Stone, 1966; Emanuel, 1979, 1982). 
It should be noted that
it is impossible in the strict sense to say
which instability is the origin of the symmetric cells, 
since the symmetric cells of the present study are
disturbances in the statistically equilibrium states
of a long time integration. 
Stability of a prescribed initial field
and time evolution of the field
are mainly concerned in the instability problems. 
When one relates disturbances in the equilibrium states
to some kind of instability, 
one should appropriately choose a reference state
that is thought to be an initial state for the instability. 
A reference state is sometimes defined
by an idealistic balanced state of forcings
prescribed to the systems under consideration. 
An alternative definition of a reference state 
is given by a simple average (in time or in space)
of the equilibrium states. 
In the following, 
relation between the symmetric cells and instabilities is considered 
according to the two kind of the reference states. 
\vspace{5mm}

Forcings imposed on the present system are
the radiative cooling below the tropopause ($\sigma =$ 0.1)
and the sensible and the latent heats from the surface. 
If any motion is suppressed, 
a balanced state of the forcings is 
in principle statically unstable. 
\footnotemark
\footnotetext{
  Radiative equilibrium will be established in each latitude
  if the radiation transfer models are used. 
  In the present study, however, 
  only the body cooling is simply used
  so that there may be a balance 
  between the radiative cooling and the thermal diffusion. 
}
Convective motions will be produced
to release the unstable stratification. 
In this respect, 
it is natural to consider
the symmetric cells to be the circulations
derived from the ``convective instability''(CI)
rather than SI. 
\vspace{5mm}

As for the reference state as an average of the equilibrium states, 
a time-mean or space-mean state has, 
in general, no relation with the initial state for the instability;
the initial state will be lost when instability occurs, 
and may be no longer similar to 
the state of the full development of the instability. 
It is well known, however, 
that the equilibrium state is sometimes
very close to the critical state for the instability, 
or becomes a slightly unstable state, 
as represented by ``convective adjustment''
and ``baroclinic adjustment''. 
If this idea is also true for the present study, 
one may infer the origin of the symmetric cells
by examining a reference state
as a time-average of the equilibrium states. 
\vspace{5mm}

Fig. 14 shows
the time-mean fields over the days 50 -- 100
in the case of Fig. 9 ($T_{A} =$ 320 K, $T_{B} =$ 300 K):
relationships between 
the normalized angular momentum $l = (u - fy)/y_{0}$
and (a) potential temperature $\theta$, 
(b) equivalent potential temperature $\theta_{e}$
and (c) saturation equivalent potential temperature $\theta_{e}^{*}$. 
Since the contours of $l$ are inclined more vertically than 
those of $\theta$
(i.e., the Richardson number is much greater than one), 
it is stable for the ``dry'' SI
(Stone, 1966). 
As for the ``moist'' SI, 
Bennetts and Hoskins (1979) discussed that
moist conditional SI is preferable
if the contours of $\theta_{e}$ are more inclined than those of $l$. 
Fig. 14(b) shows that it is stable in the upper layers. 
In the lower layers, however, 
it is conditionally ``convectively'' unstable
rather than conditionally symmetrically unstable. 
In this respect, therefore, 
the symmetric cells seem to be ascribable to CI. 
\vspace{5mm}

As seen above, 
the reference states of the present study 
is preferable for CI rather than SI, 
then it seems difficult to relate
the properties of the symmetric cells to SI. 
In the moist situation, however, 
a view of the moist SI may still be applied to the symmetric cells, 
if one isolates circulations associated with cumulus heating
from the effect of moist process on stratification. 
Emanuel (1982) considered moist SI as CISK
in a conditionally statically unstable state with a baroclinic flow. 
He pointed out that
a symmetric mode which propagates toward warmer regions exists. 
This result is consistent with the present study, 
in which the symmetric cells are always propagating
toward the warmer regions. 
It should be reminded that, in the present case, 
it is not clear whether moist circulations can be separately considered
from its effect on the stratification as in Emanuel, 
since the stratification of the system
is maintained by the symmetric cells themselves. 
\vspace{5mm}

As for the scale of the symmetric cells, 
it does not seem to be relevant to
the scale predicted by the theories of SI. 
According to the dry SI theory, 
there are two horizontal scales:
the horizontal wavelength
and the horizontal projection of the isentropes
based on the layer depth 
(Emanuel, 1979; Thorpe and Rotunno, 1989). 
The wavelength of the dry SI is controlled 
by frictional process particularly in the inviscid limit, 
but the viscosity is not important for the cell scale of the present study, 
although it may control the scale of the upward branch. 
The second scale, 
the horizontal projection of the isentropes, is given by
\begin{equation}
L_{\theta} 
= 2H \frac{\left| \DP{\theta}{z} \right|}{\left| \DP{\theta}{y} \right|}, 
                                                    \label{eq:L_theta}
\end{equation}
where $H$ is the layer depth. 
This scale seems to correspond to the wavelength of the moist SI
(Emanuel, 1982; Bennetts and Hoskins, 1979). 
Emanuel(1982) normalized horizontal length by $L_{\theta}$, 
and showed
the wavelength of the most growing mode is proportional to $L_{\theta}$
(Eq. (8) of Emanuel(1982)). 
In the numerical experiments by Bennetts and Hoskins(1979), 
the cell scale is also approximately $L_{\theta}$, 
although they did not clearly argue about the scale. 
However, the scale of the present symmetric cells
is irrelevant to $L_{\theta}$;
the scale is rather related to the slope of the angular momentum:
\begin{eqnarray}
L &=& \frac{2gH\Delta T_{s}}{f^{2}y_{0}T_{0}}
   = 2H \frac{\DP{U}{z}}{f}
= 2H \frac{\left| \DP{l}{z} \right|}{\left| \DP{l}{y} \right|}, 
                                                    \label{eq:L_m}
\end{eqnarray}
where the effect of stability is neglected as in Eq. (\ref{eq:L_delT_f}). 
Fig. 14(a) shows that
this scale is much smaller than $L_{\theta}$. 
\vspace{5mm}

  \newpage
%
\section{Summary and further comments}

  The symmetric cells which exist in the two-dimensional model
  are examined
  through the dependency on the surface temperature gradient. 
  In the case of no surface temperature gradient
  ($\Delta T_{s} =$ 0 ),
  the circulations of the symmetric cells oscillate.
  The oscillation period is given
  by the convective time $\tau_{D}$,
  and the cell scale by the Rossby deformation radius.
  Advection of the angular momentum
  by the convective motion of the cells
  is essential for the oscillation.
  In the case of a surface temperature gradient
  ($\Delta T_{s} \neq$ 0 ),
  the symmetric cells propagate toward warmer regions. 
  Each symmetric cell consists of
  a cell in the warmer side of the precipitating region
  and a cell in the colder side. 
  The cell shapes are different between the two types of the cells;
  the width of the cells in the colder side
  is broader as in the upper layers, 
  whereas the width in the warmer side is smaller as in the upper layers.
  Successive cells are sloping
  and vertically in contact with each other,
  so that the angular momentum is vertically transferred
  by diffusive process
  from the upper cell ( in the warmer region ) 
  to the lower cell ( in the colder region ). 
  \vspace{5mm}

  The Hadley cell in the axisymmetric model can be viewed
  as an extreme case of the symmetric cells;
  it is the only symmetric cell that does not move. 
  It is because a pair of the Hadley cells of both hemisphere
  is symmetric about the equator
  and the upward motion of the Hadley cells
  is located at the warmest region (the equator).
  The structure of the symmetric cells is
  very similar to that of the symmetric Hadley cell;
  the cell width of colder regions ( higher latitudes )
  is broader as in the upper layers.
  The scale of the symmetric cells is 
  also determined by a similar mechanism
  to that of the Hadley cell.
  It is determined by the relation between the surface temperature
  and the vertical mean temperature
  which is in the thermal wind balance
  with the angular momentum flow in the upper layers.
  In the Hadley case,
  the latitudinal dependency of the Coriolis parameter
  ( the equatorial $\beta$ effect )
  must be considered in the estimation of the cell width
  (S94, Fig. 13).
  \vspace{5mm}

  To show how the Hadley cell and the symmetric cells are
  related with each other,
  we calculate the motion of the position of the symmetric cells
  in the global condition. 
  If the surface temperature is given by Eq. (\ref{eq:Ts}), 
  the substitution of a local Coriolis parameter
  and a surface temperature gradient
  into Eq. (\ref{eq:L_delT_f}) yields
  \begin{eqnarray}
  L &=& \frac{gH}{2 \Omega_{0}^{2} \sin^{2} \varphi T_{0}}
        \left| \frac{1}{a} \DP{T_{s}}{\varphi} \right|
    = \frac{gH \Delta T_{s}}
             {\Omega_{0}^{2} a T_{0}} \cot \varphi
                                                    \nonumber \\
    &\equiv& R a \cot \varphi, 
                                                    \label{eq:L_global}
  \end{eqnarray}
  where $\Delta T_{s} \equiv T_{B} - T_{A}$
  is the temperature difference from the pole to the equator,
  $a$ is the radius of the earth, 
  and $R \equiv gH \Delta T_{s}/(\Omega_{0}^{2} a^{2} T_{0})$.
  If the latitude of a precipitating region is denoted by $\varphi(t)$,
  the propagation speed is estimated by
  \begin{equation}
  \DD{\varphi}{t} = - \frac{L}{a \tau_{D}}
                  = - \frac{R}{\tau_{D}} \cot \varphi.
                                                    \label{eq:dphidt}
  \end{equation}
  Integration for time
  under the initial condition $\varphi(0) = \varphi_{0}$
  yields
  \begin{equation}
  \varphi(t) = \cos^{-1}
               \left( \cos \varphi_{0} \cdot e^{\frac{R t}{\tau_{D}}} \right). 
                                                    \label{eq:cell_pos}
  \end{equation}
  Fig. 15 shows the calculated paths of the cells,
  where $\tau_{D} =$ 20 days and $R =$ 0.068 are used.
  In this figure,
  the paths in the higher latitudes than 60$^{\circ}$ are omitted.
  Fig. 15 shows that Eq. (\ref{eq:cell_pos}) reproduces
  a similar pattern to that of Fig. 2(c).
  According to Eq. (46) of S94,
  the Hadley scale is given by
  $\varphi_{H} = \sin^{-1}(\sqrt{2R/(1+2R)}) = 20.2^{\circ}$.
  It is comparable to
  the scale of the symmetric cells
  at the moment they reach the equator.
  \vspace{5mm}

  As for the results for $\Delta T_{s} =$ 0
  as shown by Fig. 4, 
  the concentration of the precipitation to the equator becomes less clear
  as the surface temperature gradient becomes smaller,
  while a cellar structure still exists in the mid-latitudes. 
  This difference may be attributed to the effect of rotation;
  moist convective motions are organized into a cellar structure
  if there is sufficient rotation.
  The discussion by Schneider(1977) should be reconsidered
  in terms of these results. 
  Schneider estimated the Hadley width
  on the assumption that a reference temperature is horizontally uniform. 
  This assumptions corresponds to the case $\Delta T_{s} =$ 0
  of the present study. 
  However, the Hadley cell no longer exists in this case
  and Schneider's estimation is not applicable. 
  Of course, it is too early to say that
  the Hadley cell in a three-dimensional model 
  will show the same behavior as
  in the present two-dimensional models
  in the case of $\Delta T_{s} =$ 0. 
  It has been reported 
  that the concentration of the precipitation to the equator
  is sensitive to cumulus parameterization, 
  in particular, 
  when the surface temperature gradient is small near the equator
  (Hess et al., 1993; Numaguti, 1993). 
  \vspace{5mm}

  It is not clear how the symmetric cells are relevant
  to the phenomena in the observed atmospheres. 
  The latitudinal motion of the cumulus bands such as ITCZ
  may have a common mechanism of the movement of the symmetric cells. 
  Goswami and Shukla(1984) simulated a northward propagation
  of the Indian monsoon by using an axisymmetric GCM. 
  In their calculation, 
  there exists a disturbance propagating toward the warmer region
  with period around 15 days, 
  and this disturbance has many similarities
  to the observed fluctuation of the Indian monsoon. 
  The structure of their disturbance is confined to the lower atmosphere
  and they pointed out the interaction with the land surface
  through the evaporation plays a key role. 
  These properties are not seen in the present study, 
  so it is not clear their disturbance and 
  the symmetric cells of the present study have the same origin. 
  \vspace{5mm}

  Although the symmetric cells are convective modes, 
  they should be re-examined
  in terms of the symmetric instability
  (e.g., Stone, 1966; Emanuel, 1979, 1982). 
  In particular, 
  Emanuel(1982) pointed out that
  a symmetric mode which propagates toward warmer regions 
  exists in a shear flow of moist atmosphere. 
  His results are consistent with the present study. 
  It will be also interesting to study the relationship 
  with the motion of the spiral bands of typhoon
  or with the phenomena in the planetary atmospheres
  such as the striped pattern of Jupiter.
  \vspace{5mm}

  The symmetric cells are also expected to exist 
  in the dry atmosphere or in the Boussinesq fluid, 
  although all the experiments in the present study are
  for the moist atmosphere. 
  They will oscillate under the uniform surface temperature condition, 
  and will propagate toward warmer regions
  if there is a surface temperature gradient, 
  since these are the consequences of the conservation
  of the angular momentum. 
  The difference will be in their horizontal scale;
  horizontal viscosity will play a major role in the scale of the cells
  in the dry atmosphere or in the Boussinesq fluid, 
  as in the case of the B\'{e}nard convection. 
  The symmetric cells should be discussed
  in a more general context, 
  in which they are viewed as a convective motion
  on a rotating plane
  with a rigid boundary below and a slip boundary above. 
  It should be also examined
  how the symmetric cells are modified 
  in a large-scale three-dimensional field, 
  particularly in the situation
  where both vertical convection and baroclinic waves co-exist
  in a vertically unstable basic stratification.
  \vspace{5mm}

  \newpage
%
{\it Acknowledgments. }
\label{sec:thanks}
\vspace{3ex}

  The comments by two referees are very helpful
  for the improvement of the manuscript. 
  This work was supported 
  by the Center for the Climate System Research, University of Tokyo. 
  Numerical calculations were performed
  with HITAC S-3800 and M-880
  at the Computer Center of the University of Tokyo, 
  and NEC SX-3 at the National Institute for Environmental Studies. 
  Figures were produced by GFD-DENNOU libraries
  and GTOOL3 by Dr. Numaguti. 
  \vspace{3ex}

  \newpage
%
\begin{center}
{\large APPENDIX} \\
{\bf \large Model formulation}
\end{center}

The model is two-dimensional
with a lateral direction and a vertical direction. 
The primitive equations
with a $\sigma$-coordinate are used. 
The global model, that is, 
a two-dimensional axisymmetric model with a spherical coordinate
is described in Sec 2 of S94. 
Formulation of the $f$-plane model ( with the Cartesian coordinate )
is almost similar to that of the global model. 
The symbols are summarized in Table. 1. 
\vspace{5mm}

Let the direction perpendicular to the two-dimensional plane
be denoted by $x$
(longitude, in the spherical case), 
the lateral direction by $y$, 
the vertical direction by $z$, 
and the corresponding metrics by 
$h_{1}$, $h_{2}$, and $h_{3}$, respectively. 
We also define $\sqrt{g} = h_{1} h_{2} h_{3}$. 
The lateral coordinate is bounded by $Y_{0} \leq y \leq Y_{1}$, 
where $Y_{0} = 0$ and $Y_{1} = 1$
if $y$ is non-dimensionalized by 
the length of the lateral domain $y_{0}$
in the Cartesian coordinate system
( Only in this appendix, $y$ is non-dimensional). 
In the spherical coordinate system, 
$y = \sin \varphi$ ($\varphi$ is the latitude) is used, 
and hence $Y_{0} = -1$ and $Y_{1} = 1$. 
The metrics and the angular momentum $l$
are expressed, in the Cartesian coordinate, by
\begin{eqnarray*}
( h_{1}, h_{2}, h_{3} ) &=& ( 1, y_{0}, 1 ), 
                                                \\
l &=& u - f y y_{0}. 
\end{eqnarray*}
whereas, in the spherical coordinate, by
\begin{eqnarray*}
( h_{1}, h_{2}, h_{3} ) &=& \left( a \sqrt{ 1 - y^{2}},
                                 \frac{a}{\sqrt{1 - y^{2}}}, 1 \right), 
                                                \\
l &=& u h_{1} + \Omega h_{1}^{2}, 
\end{eqnarray*}
where $f = 2 \Omega$ is the Coriolis parameter, 
$\Omega$ is the rotation velocity, 
and $a$ is the radius of the earth. 
\vspace{5mm}

The surface pressure $p_{s}$, 
the angular momentum $l$,
the lateral velocity $v$, 
the moist enthalpy $h$, 
and the specific humidity $q$
are predicted variables in this system. 
The choice of $l$ and $h$ as the predicted variables
automatically guarantees
the conservations of the angular momentum and the moist enthalpy. 
The basic equations are expressed as follows:
\begin{eqnarray}
\partial_{t} p_{s} &=& - \frac{1}{\sqrt{g}} 
                     \partial_{y} ( h_{1} p_{s} v )
                   - \partial_{\sigma} ( p_{s} \dot{\sigma} ), 
                                              \label{eq:model_ps}\\
\partial_{t} ( p_{s} l ) &=& - \frac{1}{\sqrt{g}} 
                     \partial_{y} ( h_{1} p_{s} v l )
                   - \partial_{\sigma} ( p_{s} \dot{\sigma} l )
                   + h_{1} p_{s} f_{x}, 
                                              \label{eq:model_l} \\
\partial_{t} ( p_{s} v ) &=& - \frac{1}{\sqrt{g}} 
                     \partial_{y} ( h_{1} p_{s} v^{2} )
                   - \partial_{\sigma} ( p_{s} \dot{\sigma} v )
                   - \frac{1}{h_{2}} 
                     \left[ \frac{p}{\rho}
                       \partial_{y} p_{s}
                     + p_{s} \partial_{y} \Phi
                     \right]
                                                               \nonumber \\
&&
                   + p_{s} F_{C}
                   + p_{s} f_{y}, 
                                              \label{eq:model_v}\\
0 &=& - p_{s} - \rho \partial_{\sigma} \Phi, 
                                              \label{eq:model_seiriki}\\
\partial_{t} ( p_{s} h ) &=& - \frac{1}{\sqrt{g}} 
                     \partial_{y} ( h_{1} p_{s} v h )
                   - \partial_{\sigma} ( p_{s} \dot{\sigma} h )
                                                               \nonumber \\
&&
                   + \frac{p_{s}}{\rho} \DD{}{t} p
                   + \frac{p_{s}}{\rho} Q^{dif}
                   + \frac{p_{s}}{\rho} Q^{rad}
                   + \frac{p_{s}}{\rho} Q^{fric}, 
                                              \label{eq:model_h} \\
\partial_{t} ( p_{s} q ) &=& - \frac{1}{\sqrt{g}} 
                     \partial_{y} ( h_{1} p_{s} v q )
                   - \partial_{\sigma} ( p_{s} \dot{\sigma} q )
                   + \frac{p_{s}}{\rho} S^{dif}
                   - \frac{p_{s}}{\rho} S^{prc}, 
                                              \label{eq:model_q}
\end{eqnarray}
where $\Phi = gz$ is the gravitational potential
and $F_{C}$ is the centrifugal force, 
which is given, in the spherical coordinate, by 
\begin{equation}
F_{C} =  - \frac{1}{h_{2}} \partial_{y} \Phi_{C}
                   - l^{2} \frac{1}{h_{2}} 
                     \partial_{y} \frac{1}{2 h_{1}^{2}}, 
                                                    \label{eq:FC}
\end{equation}
where $\Phi_{C} = - \Omega^{2} h_{1}^{2}/2$ 
is the centrifugal potential. 
In the Cartesian coordinate, 
it corresponds to the Coriolis force $F_{C} = f u$. 
The vertical $\sigma$-velocity $\dot{\sigma}$
is given by the integration from $\sigma$ to $\sigma =$ 1 
of Eq. (\ref{eq:model_ps}):
\begin{eqnarray}
\dot{\sigma}
&=& \frac{1}{p_{s}}
    \left[ ( 1 - \sigma ) \partial_{t} p_{s}
           + \frac{1}{\sqrt{g}} 
             \partial_{y}
                  \left( h_{1} p_{s} \int_{\sigma}^{1} v \,d\sigma
                  \right)
    \right]. 
                                              \label{eq:model_sigdot}
\end{eqnarray}
\vspace{5mm}

The diffusion terms
$f_{x}$, $f_{y}$, $Q^{fric}$, $Q^{dif}$, and $S^{dif}$
in the basic equations are given by the same form as in section 2 of S94. 
$S^{prc}$ of the humidity equation
is a term representing the precipitation, 
which is calculated as a removal of a liquid part
of the humidity after each of time integration
(this process is referred as the ``large-scale condensation''). 
The radiative cooling $Q^{rad}/\rho$ is specified as a constant value
in the $f$-plane model. 
Boundary conditions are
$\dot{\sigma} = 0$ at $\sigma =$ 0, 1, 
and $v = 0$ at $y = Y_{0}$, $Y_{1}$. 
In the spherical coordinate, 
although $l = 0$ at the poles ($y = Y_{0}, Y_{1}$), 
the angular velocity $l/h_{1}^{2}$ does not generally vanish. 
The fluxes at the boundaries at $y = Y_{0}, Y_{1}$, and $\sigma = 0$
are zero, 
while those at $\sigma = 1$ are given by bulk formulae. 
\vspace{5mm}

The grid points are defined as follows. 
As shown by Fig. 16, 
the lateral boundary $y = Y_{0}$ is denoted by $j = $ 1/2, 
and $y = Y_{1}$ by $j = J + 1/2$, 
while the vertical boundary $\sigma = 1$ is denoted by $k = 1/2$
and $\sigma = 0$ by $k = K + 1/2$. 
The grid points are defined at both half-integer $j+1/2$ ( $k+1/2$ )
and integer $j$ ($k$). 
The half-integer point 
which is placed between $j$ and $j+1$ ( $k$ and $k+1$ )
is denoted by $j + 1/2$ ( $k + 1/2$ ). 
The number of the half-integer points is $J+1$ in the lateral direction, 
while it is $K+1$ in the vertical direction. 
$J$ is even in the case of the spherical coordinate system. 
The grid intervals of the lateral direction are equal
both for integer and half-integer points. 
As for the vertical direction, 
the half-integer points are equally placed, 
while the integer points are based on Arakawa and Suarez(1983):
\begin{eqnarray}
y_{j-\frac{1}{2}} &=& Y_{0} + ( Y_{1} - Y_{0} ) \frac{j-1}{J};
\mbox{\hspace{20mm}} ( j = 1, ..., J+1 ), 
                                              \label{eq:model_y_half} \\
y_{j}
&=& \frac{1}{2} \left( y_{j+\frac{1}{2}} + y_{j-\frac{1}{2}}
              \right);
\mbox{\hspace{25mm}} ( j = 1, ..., J ). 
                                              \label{eq:model_y_int} \\
\sigma_{k-\frac{1}{2}} &=& 1 - \frac{k-1}{K}; 
  \mbox{\hspace{40mm}} ( k = 1, ..., K+1 ). 
                                          \label{eq:model_sigma_half} \\
\sigma_{k}
&=& \left( \frac{1}{\kappa}
         \frac{   \sigma_{k - \frac{1}{2}}^{\kappa+1}
                - \sigma_{k + \frac{1}{2}}^{\kappa+1} }
              {   \sigma_{k - \frac{1}{2}}
                - \sigma_{k + \frac{1}{2}} }
  \right)^{\frac{1}{\kappa}};
  \mbox{\hspace{20mm}} ( k = 1, ..., K ), 
                                            \label{eq:model_sigma_int}
\end{eqnarray}
where $\kappa = R_{d} / C_{p}$. 
\vspace{5mm}

The variables are defined at the following points:
\begin{displaymath}
\begin{array}{lll}
p_{s} & p_{s}{}_{j-\frac{1}{2}}; & ( j = 1,...,J+1 )\\  
l     & l_{j-\frac{1}{2},k};   & ( j = 1,...,J+1; k = 1, ..., K ) \\  
v     & v_{j,k};         & ( j = 1,...,J;   k = 1, ..., K ) \\  
h     & h_{j-\frac{1}{2},k};   & ( j = 1,...,J+1; k = 1, ..., K ) \\  
q     & q_{j-\frac{1}{2},k};   & ( j = 1,...,J+1; k = 1, ..., K ) \\
\dot{\sigma} & \dot{\sigma}_{j-\frac{1}{2},k-\frac{1}{2}};
                               & ( j = 1,...,J+1; k = 1, ..., K+1 )
\end{array}
\end{displaymath}
Vertical discretization by 
Arakawa and Lamb(1977) and Arakawa and Suarez(1983)
is used so as to conserve total kinetic energy. 
The angular momentum and the equations of motion are discretized 
so as to conserve total kinetic energy(Satoh, 1994b).

  \newpage
%
\begin{center}
{\bf \large REFERENCES}
\end{center}

\begin{description}

\item Arakawa, A. and V. R. Lamb, 1977 :
      Computational design of the basic dynamical processes of the UCLA
      general circulation model, 
      {\it Methods in Computational Physics}, 
      {\bf 17}, 173-265.

\item Arakawa, A and M. J. Suarez, 1983 :
      Vertical differencing of the primitive equations 
      in sigma coordinates,
      {\it Mon. Weath. Rev.}, {\bf 111}, 34-45. 
 
\item Bennetts, D. A. and B. J. Hoskins, 1979:
      Conditional symmetric instability --
      a possible explanation for frontal rainbands. 
      {\it Quart. J. R. Met. Soc.}, {\bf 105}, 945-962. 

\item Bjerknes, V., 1937 :
      Application of line integral theorems 
      to the hydrodynamic of terrestrial and cosmic vortices, 
      {\it Astrophysica Norv.}, {\bf 11}, 263-339.

\item Bretherton, C.S., 1987 : 
      A theory for nonpredicting moist convection between two parallel. 
      Part I: Thermodynamics and ``linear'' solutions. 
      {\it J. Atmos. Sci.}, {\bf 44}, 1809-1827. 

\item Eliassen, A., 1951 :
      Slow thermally or frictionally controlled 
      meridional circulation in a circular vortex, 
      {\it Astrophysica Norv.}, {\bf 5}, 19-60. 

\item Emanuel, K. A., 1979 :
      Inertial instability and mesoscale convective systems. Part I :
      Linear theory of inertial instability in rotating viscous fluids, 
      {\it J. Atoms. Sci.}, {\bf 36}, 2425-2449.

\item Emanuel, K. A., 1982 :
      Inertial instability and mesoscale convective systems. Part II :
      Symmetric CISK in a baroclinic flow, 
      {\it J. Atmos. Sci.}, {\bf 39}, 1080-1097. 

\item Emanuel, K. A., 1983 :
      Elementary aspects of the interaction between cumulus convection
      and the large-scale environment. 
      In
      {\it Mesoscale Meteorology -- Theories, Observations and Models}
      ( Eds. Lilly, D.K. and Gal-Chen, T. ), 
      Reidel, 551-575. 

\item Ferrel, W., 1856 :
      An essay on the winds and currents of ocean, 
      {\it Nashville Journal of Medicine and Surgery}, {\bf 11}. 

\item Gill, A. E., 1982 :
      {\it Atmosphere-Ocean Dynamics.} 
      Academic Press, 662pp. 

\item Goswami, B.N. and J. Shulka, 1984 : 
      Quasi-periodic oscillations in a symmetric general circulation model. 
      {\it J. Atmos. Sci.}, {\bf 41}, 20-37. 

\item Held, I. M. and A. Y. Hou, 1980 :
      Nonlinear axially symmetric circulations 
      in a nearly inviscid atmosphere,
      {\it J. Atmos. Sci.}, {\bf 37}, 515-533.

\item Hess, P. G., D. S. Battisti and P. J. Rasch, 1993 :
      Maintenance of the intertropical convergence zones
      and the large-scale tropical circulation on a water-covered earth, 
      {\it J. Atmos. Sci.}, {\bf 50}, 691-713. 

\item Lorenz, E., 1967 :
      The nature and theory of the general circulation of the 
      atmosphere, 
      {\it World Meteorological Organization }, 161pp.

\item Numaguti, A., 1993 : 
      Dynamics and energy balance of the Hadley circulation
      and the tropical precipitation zones:significance of
      the distribution of evaporation, 
      {\it J.Atmos.Sci.}, {\bf 50}, 1874-1887. 

\item Plumb, A. and A. Y. Hou, 1992 : 
      The response of a zonally-symmetric atmosphere
      to subtropical thermal forcing : threshold behavior. 
      {\it J.Atmos.Sci.}, {\bf 49}, 1790-1799. 

\item Satoh, M., 1994a :
      Hadley circulations in radiative-convective equilibrium
      in an axially symmetric atmosphere, 
      {\it J.Atmos. Sci.}, {\bf 51}, 1947-1968. 

\item Satoh, M., 1994b :
      Hadley circulations in radiative-convective equilibrium
      in an axially symmetric atmosphere (in Japanese), 
      Ph.D thesis, Univ. of Tokyo, 178pp. 

\item Schneider, E. K., 1977 :
      Axially symmetric steady-state models of the basic state 
      for instability and climate studies. II. Nonlinear calculations, 
      {\it J. Atmos. Sci.}, {\bf 34}, 280-296. 

\item Stone, P. H., 1966 :
      On non-geostrophic baroclinic stability, 
      {\it J. Atmos. Sci.}, {\bf 23}, 390-400.

\item Thomson, J., 1892 :
      On the grand currents of atmospheric circulation, 
      {\it Phil. Trans. Roy. Soc., A}, {\bf 183}, 653-684. 

\item Thorpe, A. J. and R. Rotunno, 1989 : 
      Nonlinear aspects of symmetric instability. 
      {\it J. Atmos. Sci.}, {\bf 46}, 1285-1299. 

\end{description}

  \newpage
\begin{center}
{\bf \large Figure Captions}
\end{center}
\label{sec:figcap}

\begin{description}
  \item[Fig. 1.] 
           Convergence of the angular momentum. 
           The standard experiment of the global model. 
           (a) total convergence
               $- \overline{ \Ddiv ( lv )}$, 
           (b) steady component $- \Ddiv ( \overline{l} \overline{v} )$, 
           and (c) unsteady component $- \Ddiv ( \overline{l' v'} )$. 
              The contour intervals are
              (a) 100 m$^{2}$/s$^{2}$, 
              and (b), (c) 60 m$^{2}$/s$^{2}$. 
  \item[Fig. 2.] 
           Time variation of the meridional distribution of the precipitation. 
           Dependencies on the pole-to-equator temperature difference:
           (a) $\Delta T_{s} = 0$ K, 
           (b) $\Delta T_{s} = 20$ K, 
           (c) $\Delta T_{s} = 40$ K (the standard experiment), 
           and (d) $\Delta T_{s} = 80$ K. 
           The contour intervals are 5 $\times$ 10$^{-5}$ kg/m$^{2}$ s. 
  \item[Fig. 3.] 
           Time variation of the meridional distribution of the precipitation. 
           Dependencies on the coefficient of the viscosity:
           (a) $\nu = \nu_{s} = 1$ m$^{2}$/s, 
           (b) $\nu = \nu_{s} = 25$ m$^{2}$/s, 
           where $\nu$ is the viscosity in the free atmosphere
           and $\nu_{s}$ is the viscosity at the surface. 
           The contour intervals are 5 $\times$ 10$^{-5}$ kg/m$^{2}$ s. 
  \item[Fig. 4.] 
           Time variation of the lateral distribution of the precipitation
           in the $f$-plane model. 
           The surface temperature is linearly distributed
           from $T_{A}$ at $y/y_{0} = 0$
           to $T_{B}$ at $y/y_{0} = 1$:
           (a) $T_{A} = T_{B} = 300$ K, 
           (b) $T_{A} = 305$ K, $T_{B} = 295$ K, 
           (c) $T_{A} = 310$ K, $T_{B} = 290$ K, 
           and (d) $T_{A} = 320$ K, $T_{B} = 280$ K. 
           The contour intervals are 2 $\times$ 10$^{-4}$ kg/m$^{2}$ s. 

  \item[Fig. 5.] 
           Time variation of the zonal winds 
           ( winds perpendicular to the model domain )
           at the surface. 
           Surface temperature is uniform : $T_{A} = T_{B} = 300$ K. 
           The contour intervals are 2 m/s. 
  \item[Fig. 6.] 
           Distribution of the zonal winds
           ( winds perpendicular to the figures ).
           Surface temperature is uniform : $T_{A} = T_{B} = 300$ K. 
           (a) 10 days average for the days 46 -- 55, 
           and
           (b) 10 days average for the days 56 -- 65. 
           The contour intervals are 5 m/s. 
  \item[Fig. 7.] 
           As in Fig. 6 but for the temperature difference
           from the horizontal average. 
           The contour intervals are 1 K.
  \item[Fig. 8.] 
           Time variation of the temperature lapse rate
           ($- dT/dz$) at $\sigma=0.9$. 
           Surface temperature is uniform : $T_{A} = T_{B} = 300$ K. 
           The contour intervals are 2 K/km.
           The hatched area indicates the values larger than 8 K/km. 
  \item[Fig. 9.] 
           Time variation of the lateral distribution of the precipitation
           in the $f$-plane model
           for $T_{A} =$ 320 K and $T_{B} =$ 300 K. 
           The contour intervals are 2 $\times$ 10$^{-4}$ kg/m$^{2}$ s. 
           A $\rightarrow$ B denotes a reference line for the composite maps. 
  \item[Fig. 10.] 
           Composite maps averaged with 
           the propagation of 
           the symmetric cells
           from $y/y_{0}=0.68$ at the day 51
           to $y/y_{0}=0.45$ at the day 100
           (shown by A $\rightarrow$ B in Fig. 9). 
           The center of the symmetric cell is placed at $y/y_{0} = 0.5$. 
           (a) zonal winds(winds perpendicular to the figure)
               with the contour interval of 5 m/s
           and (b) stream functions
               with the contour interval of 1,000 kg/m s.
               (Stream functions are those per unit length
                perpendicular to the figure. )
  \item[Fig. 11.] 
           Temperature (solid lines) and 
           normalized angular momentum
           $(u - fy)/fy_{0}$ (broken lines)
           at the day 100
           in the case of $T_{A} =$ 320 K and $T_{B} =$ 300 K. 
           The contour intervals are 5K for the temperature
           and 0.02 for the normalized angular momentum. 

  \item[Fig. 12.] 
           The scales of the symmetric cells. 
           (a) scale in the case of $\Delta T_{s} =$ 0 : $R_{N}$, 
           (b) scale in the case of $\Delta T_{s} \neq$ 0
               when the effect of static stability is not included
               : $R_{T}$, 
           and 
           (c) scale in the case of $\Delta T_{s} \neq$ 0
               when the effect of static stability is included
               : the warmer side $-L_{-}$
                 and the colder side $L_{+}$. 
           Thick allows indicate the precipitating region. 
           The curves are the temperature difference
           from the precipitating region
           in the middle of the troposphere $\delta T$, 
           and the sloping lines are that of the surface temperature
           $\delta T_{s}$. 
           The three vertical segments of lines shown by $dT_{0}$
           in (a) and (c) are the same temperature difference
           $(g/C_{p} - \Gamma_{m}) H/2$. 
           In (b), $dT_{1} = \Delta T_{s} R_{T}/y_{0}$, 
           and, in (c),  $dT_{2} = \Delta T_{s} L_{+}/y_{0}$. 
  \item[Fig. 13.] 
  Relationship between surface zonal winds
  and motion of a parcel
  which starts from the origin $y =$ 0
  shown by a thick arrow. 
  $y =$ 0 is the position of precipitation
  at the time when the parcel starts. 
  If the angular momentum is conserved along the motion of the parcel, 
  zonal wind becomes easterly if the parcel comes down to the negative $y$, 
  whereas it becomes westerly if the parcel comes down to the positive $y$. 
  Zonal wind is zero only if the parcel returns to
  the starting position $y =$ 0. 
  In order for the surface wind to be easterly, 
  the precipitation must move toward the negative direction
  during the cycle of the parcel. 
  \item[Fig. 14.] 
           Comparison of time averaged distributions of
           (a) potential temperature $\theta$, 
           (b) equivalent potential temperature $\theta_{e}$
           and 
           (c) saturation equivalent potential temperature $\theta_{e}^{*}$
           with distribution of normalized angular momentum $(u - fy)/fy_{0}$. 
           The normalized angular momentum is the same in (a)-(c), 
           and is shown by broken lines with the contour interval of 0.02. 
           The potential temperatures, 
           $\theta$, $\theta_{e}$ and $\theta_{e}^{*}$, 
           are shown by solid lines with the contour interval of 5K. 
           These are 50 days average for the days 51-100
           in the case of $T_{A} =$ 320 K and $T_{B} =$ 300 K. 
  \item[Fig. 15.] 
           Calculated patterns of the symmetric cells propagation
           by Eq. (\protect\ref{eq:cell_pos}). 
  \item[Fig. 16.] 
           Arrangement of the grid points. 
\end{description}

  \newpage
%
  \vspace*{\fill}
    \begin{center}
    {\bf Table. 1} Symbols. 
    \end{center}
    \vspace{5mm}

\begin{center}
\begin{tabular}{ll}
$x$
  & coordinate perpendicular to the two-dimensional plain \\
$y$
  & lateral coordinate \\
$z$
  & altitude \\
$h_{1}$, $h_{2}$, $h_{3}$, $\sqrt{g} = h_{1} h_{2} h_{3}$
  & metrics \\
$p_{s}$
  & surface pressure \\
$p$
  & pressure \\
$\sigma = p / p_{s}$
  & vertical coordinate \\
$u$, $v$
  & velocity component \\
$l$
  & angular momentum \\
$\dot{\sigma}$
  & $\sigma$-velocity \\
$T$
  & temperatuer \\
$q$
  & specific humidity \\
$C_{p}$
  & specific heat for constant pressure \\
$R_{d}$ 
  & gas constant for dry air \\
$L$
  & latent heat \\
$h = C_{p} T + L q$
  & moist enthalpy \\
$\rho$
  & density \\
$T_{s}$
  & surface temperature \\
$\Phi = g z$
  & gravitational potentail \\
$\Phi_{c}$
  & centrifugal potential \\
$a$ 
  & radius of the earth \\
$\Omega$ 
  & rotation velocity \\
$\Omega_{0}$ 
  & rotation velocity of the earth\\
$f_{x}$, $f_{y}$
  & viscous forces \\
$Q^{dif}$ 
  & energy convergence due to thermal and moisture diffusions \\
$Q^{rad}$ 
  & energy convergence due to radiation \\
$Q^{fric}$ 
  & energy convergence due to dissipation \\
$S^{dif}$ 
  & moisture convergence due to diffusion \\
$S^{prc}$ 
  & precipitation \\
\end{tabular}
\end{center}
  \vspace*{\fill}

\end{document}